\documentclass[times, twocolumn, final]{elsarticle}

\usepackage{framed,multirow}
\usepackage{graphicx}
\usepackage{dsfont}
\usepackage{wrapfig}
\usepackage{subfig}
\usepackage{extarrows}
\usepackage{csquotes}

\usepackage{amssymb}
\usepackage{latexsym}
 \usepackage{amsmath} 
\usepackage{natbib} 
\usepackage{url}
\usepackage{xcolor}

\newcommand{\argmax}{\operatornamewithlimits{argmax}}

\begin{document}

\ifpreprint
  \setcounter{page}{1}
\else
  \setcounter{page}{1}
\fi

\begin{frontmatter}



\title{Uniform Resampling vs. Image Blur: Aliasing
Approximation via Isotropic Gaussian Filtering}

\author[label1,label2]{Suayb S. Arslan\corref{cor1}}
\author[label2]{Lukas Vogelsang}
\author[label2]{Michal Fux}
\author[label2]{Pawan Sinha}

\cortext[cor1]{Corresponding author.
  Tel.: +90-212-359-7335; e-mail: suayb.arslan@bogazici.edu.tr}

\affiliation[label1]{organization={Dept. of Computer Engineering, Bogazici University},
             city={Istanbul},
             country={Türkiye}}

\affiliation[label2]{organization={Dept. of Brain and Cognitive Sciences, Massachusetts Institute of Technology},
             state={MA},
             country={USA}}


\begin{abstract}
One of the key approximations to range simulation is downscaling the image, dictated by the natural trigonometric relationships that arise due to long-distance viewing. It is well-known that standard downsampling applied to an image without prior low-pass filtering leads to a type of signal distortion called \textit{aliasing}. In this study, we aim at modeling the distortion due to aliasing and show that a downsampled/upsampled image after an interpolation process can be very well approximated through the application of isotropic Gaussian low-pass filtering to the original image. In other words, the distortion due to aliasing can approximately be generated by low-pass filtering the image with a carefully determined cut-off frequency. We have found that the standard deviation of the isotropic Gaussian kernel $\sigma$ and the reduction factor $m$ (also called downsampling ratio) satisfy an approximate $m \approx 2 \sigma$ relationship. We provide both theoretical and practical arguments using two relatively small face datasets (Chicago DB, LRFID) as well as TinyImageNet to corroborate this empirically observed relationship.
\end{abstract}




\end{frontmatter}




\section{Introduction}
\label{intro}

Images captured by digital cameras are the result of collecting data from continuous two-dimensional sources, often represented by millions of digitally encoded color shades. The captured digital data typically contains more than what is perceivable by the human eye. Thus, it is downsampled to reduce data size and retain what is semantically meaningful. A phenomenon called \textit{aliasing} manifests itself when a digital image is sampled at a rate lower than the \textit{Nyquist rate} (a quantity related to the maximum available spatial frequency in the image), leading to unwanted distortions and various visual artifacts in the sampled image. Understanding aliasing and its relationship to downsampling and subsequent upsampling (collectively called \textit{decimation}) has been crucial in the digital signal processing and pattern recognition literature and has found various applications, such as distance viewing \citep{held2010using, groth2024comparing} and face recognition \citep{jarudi2023recognizing, arslan2023effective} to ensure accurate and faithful representation of image semantics in the resampled domain.

\begin{figure*}[t!]%
    \centering
\includegraphics[width=14.3cm]{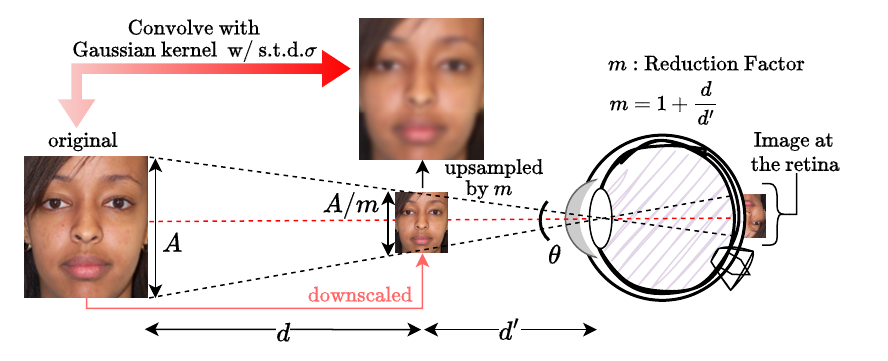}%
    \caption{Viewing an image from a distance effectively results in a downscaling of the original image. Simulating this effect through upsampling can introduce various artifacts due to aliasing. The reduction factor \( m \) can be related to viewing distances as \( m = 1 + \frac{d}{d'} \).}
    \label{distview}
    \vspace{-6mm}
\end{figure*}

\subsection{Motivation} For a uniformly sampled spatial configuration, the image field experiences effective downsampling when it is viewed from increased distances (due to a limited number of sensors or photoreceptors in the human eye), implying fewer data pixels are employed to represent the semantic meaning in the image patch. Subsequently, the image undergoes upsampling (and interpolation), ideally to its original size, resulting in a degraded version of the observed image due to aliasing or frequency mixing. As shown in Fig. \ref{distview}, due to the natural consequence of downscaling in the retina, a replacement of the original image with the upsampled version would not change the quality as seen at the retina. The relationship between the decimation ratio $m$ and the Gaussian kernel parameter $\sigma$ to approximate the degradation (through convolution) could be extremely useful for creating a frontend similar to the human visual system for comparing computational models against human task performance \citep{arslan2023distance,suayb2025githubRepo}. This way, fair comparisons can be made between humans and machines for the same task performances.

Additionally, beyond simulating distant viewing, this concept could also be relevant for replicating low-vision conditions \citep{wang2023artificial}. For example, in experiments involving individuals with acuity lower than 20/20, a reduction factor can be used to mimic visual impairment. This enables the implementation of on-screen blurring for experimental stimuli, eliminating the need for physical blur goggles and streamlining control experiments in low-vision research.

\subsection{Research Question}  To quantify the degradation required, either due to long distance viewing conditions or low vision, preferably with a single parameter (call it $\sigma$), the \textit{Gaussian} function could serve as the evident choice for approximation (see Fig. \ref{distview}). We illustrate the practical viability of this approximation, substantiating our assertions both theoretically and empirically. These tests establish a simple yet remarkably surprising relationship between the downscaling ratio ($m$) and the Gaussian kernel parameter ($\sigma$), indicating that the manipulation of this parameter allows for an accurate simulation of degradation due mainly to aliasing. In this short note, we specifically address the following research question: 
\begin{quote}
\textit{Given an image, down-sampled and then up-sampled back to the original size using 2D-interpolation (such as \textit{B-splining or bicubic}), what is the corresponding Gaussian kernel size (or standard deviation $\sigma$ that adjusts the amount of blur\footnote{The kernel size for a Gaussian filter is typically chosen to be $2\lceil 3\sigma \rceil + 1$, ensuring that more than $99\%$ of the Gaussian's total distribution is captured within the filter window. This balances computational efficiency with the need for accurate smoothing, as the Gaussian function effectively vanishes beyond three standard deviations from the mean.}) that leads to the maximum similarity (such as minimum sample-wise mean square error (MSE) or maximum structural \textit{similarity} index (SSIM)) between the resampled version and the original image convolved with this kernel$-$low-pass filtered version of the image?}
\end{quote}

Although the formulation of this question may appear quite straightforward, tackling the issue of aliasing is more complex than it seems due to its reliance on the frequency characteristics of the input signal. As a result, a comprehensive theoretical answer might transcend a basic analysis due to the challenging and multi-faceted problem of statistical modeling of the image space \citep{van1996modelling}. Consequently, we have presented a well-justified input-agnostic approximate analysis with required details provided in the Appendix. Moreover, we have complemented our analysis with a comprehensive array of experiments for numerical validation.

\section{Theory}

Let us begin by providing a summary of known results and assumptions. Although imaging systems are 2-dimensional, we provide our justification in a 1-dimensional discrete setting as it is easier to extend our results to higher dimensions by considering each dimension separately, since the Gaussian kernels can decompose. This mathematical setup is very common in the development of imaging algorithms \citep{paxman1992joint}.

A normalized Gaussian function and its Fourier Transform ($\mathcal{F}[.]$) are given by
\begin{equation}
    {g}(x,\sigma) = \frac{1}{\sigma \sqrt{2 \pi}} e^{-\frac{x^2}{2\sigma^2}} 
     \xLongleftrightarrow[\mathcal{F}]{} 
     G(w,\sigma) = e^{-\frac{w^2 \sigma^2}{2}} = e^{- \frac{w^2}{2\left(\frac{1}{\sigma}\right)^2}}
\end{equation}
where $w$ is frequency in radians, $\int_{-\infty}^{\infty} g(x, \sigma) dx = 1$, and $\mathcal{F}$ represents the Fourier Transform operator. Note that a Gaussian in the time (or space) domain with standard deviation $\sigma$ is equal to another Gaussian in the frequency domain with standard deviation $1/\sigma$ (reciprocity property). Since all digital data is stored discretely, the way the normalized Gaussian function is implemented is given by 
\begin{eqnarray}
    g(x_i,\sigma) = l_i(x)/\sum_j l_j(x) 
\end{eqnarray}
where $l_i(x) = \exp\left\{-x_i^2/{2\sigma^2}\right\}$. A discrete-time signal  gives rise to a periodic frequency spectrum with a normalized frequency period of $2\pi$ (radians) \citep{allen2004signal}. Note that all real signals have a conjugate symmetric frequency spectrum i.e., $G(w,\sigma) =  G^*(-w,\sigma)$ and hence symmetric magnitude.

\begin{figure}
  \centering
\includegraphics[width=7.6cm]{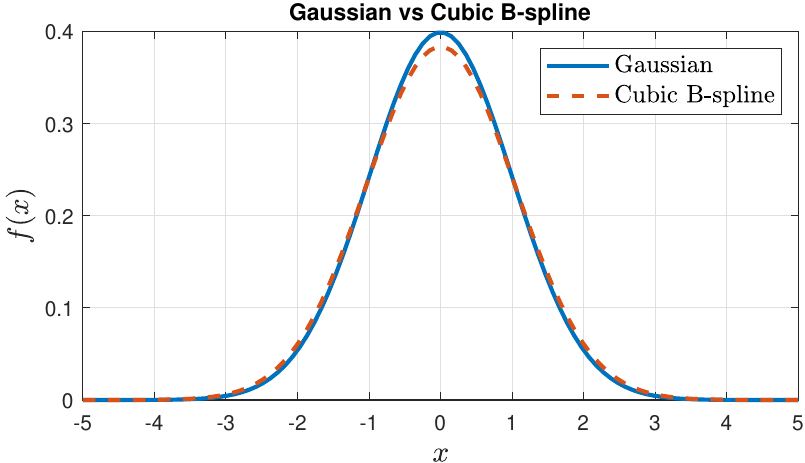}
  \caption{A comparison between the isotropic Gaussian function and the cubic B-Spline function (shown as $f(x)$ in the figure) (the Gaussian kernel is depicted using the solid line; the cubic B-spline using a dotted line).}
  \label{GaussSpline}
\end{figure}

\begin{figure*}[t!]%
    \centering
\includegraphics[width=15.2cm, height=11.1cm]{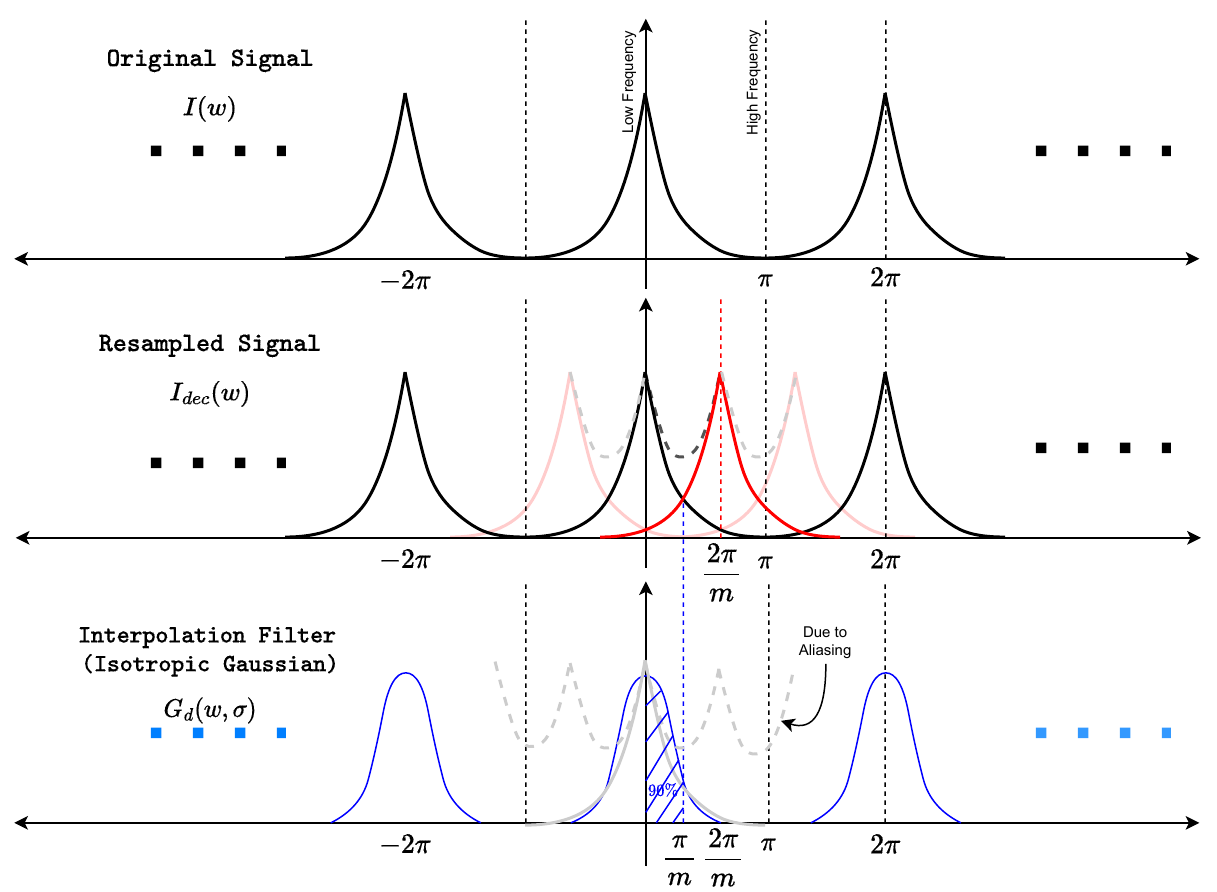}%
    \caption{An illustration of what the original and resampled (decimated) signals look like in the frequency domain. We also included the frequency response of a discrete Gaussian filter with parameter $\sigma$, superimposed on the frequency spectrum of the original signal, to illustrate how they compare to each other.}
    \label{freqSpec}
\end{figure*}


We assume that the impulse response, as well as the frequency response of the interpolation filter (\textcolor{black}{cubic B-spline -  other interpolation techniques (e.g. bicubic) yield similar results}), can be approximated by Isotropic Gaussian Filtering (low-pass filtering). An example is shown in Fig. \ref{GaussSpline} for order-12 B-splines. As can be seen, both kernels implement low-pass filtering and  share the same functional form \citep{wang1998scale}, only showing differences in point-wise slope to shape the steepness of the fall-off region in the frequency domain representations (see \citep{young30500gaussian} for more formal convergence arguments of B-splines to Gaussian function for Hilbert spaces). 

Let us suppose the continuous signal has the spectral representation $X(w) = \mathcal{F}[x(t)]$. We express our discrete (sampled) signal as well as the sampled Gaussian as follows:
\begin{eqnarray}
    I(w) &=& \sum_k X(w - 2 \pi k) \\
    G_d(w,\sigma) &=& \sum_k G(w - 2 \pi k, \sigma).
\end{eqnarray}

Reducing the size of the image by keeping the aspect ratio intact (or rescaling) corresponds to uniformly downsampling the signal (in both dimensions). Since downsampling leads to a contraction in time (or space in 2D), its frequency representation is expanded. Subsequently, if we further upsample the image (leading to an expansion in time and a contraction in frequency), the net result would be what is called \textit{decimation} of $I(w)$, which can be expressed for a reduction factor (or downscaling ratio) of $m$ as 
\begin{eqnarray}
    I_{dec}(w) = \sum_k X(w - \frac{2 \pi}{m} k).
\end{eqnarray}

\begin{figure}%
    \centering
\includegraphics[width=9cm]{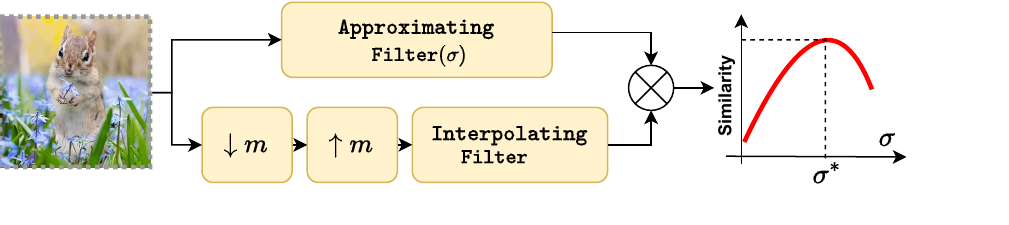}%
    \caption{The experiment setup for a given image and the two signal processing operations (a Gaussian approximating filter and a cubic B-spline interpolating filter).}
    \label{Protocol}
    \vspace{-5mm}
\end{figure}

Let us also define a similarity function that minimizes the difference (or maximizes the similarity) between the original and the estimated signals i.e., $\Psi_{sim}(I(w),\hat{I}(w))$.  The last step is to apply an interpolation filter to approximate the original signal $I(w)$ with respect to $\Psi_{sim}(.)$. Since filtering (convolution) in time domain corresponds to a multiplication in the frequency domain, we obtain the final estimation expressed in the frequency domain as
\begin{align}
    \hat{I}(w) = \sum_k G(w-2\pi k,\sigma^*) I_{dec}(w), 
\end{align}
where $\sigma^*$ is chosen to optimize the estimation accuracy (see Eqn. \eqref{eqn66}). To be able to proceed with the details of this optimization, we demonstrate visually what the representation of these signals looks like in the frequency domain in Fig. \ref{freqSpec}. The top spectrum is assumed to be the original digital signal $I(w)$. Later, this signal is decimated, which inherently creates copies of the spectra and centers them around the integer multiples of $2\pi/m$ radians. The bottom spectrum belongs to an isotropic Gaussian interpolating filter. Since we assumed the approximating and interpolating filters to be both Gaussian, their spectra look identical. The parameter of the Gaussian filter is chosen such that we maximize the following:

\small
\begin{align}
    \sigma^* = \argmax_{\sigma} \Psi_{sim}\left(\mathcal{F}^{-1}[I(w)G_d(w,\sigma)], \mathcal{F}^{-1}[I_{dec}(w)G_d(w,\sigma)]\right), \label{eqn66}
\end{align}
\normalsize
where $\mathcal{F}^{-1}$ is the inverse Fourier transform.

The overall process is illustrated in Fig. \ref{Protocol}. There are two implications of the estimation: (1) The approximating Gaussian filtering on the original image should be designed jointly with the interpolating filter (which is also assumed to be Gaussian in our scenario). (2) Since aliasing (low/high frequencies overlapping) is inevitable due to large downscaling/reduction factors, the filter choice is supposed to approximate the non-linear distortion due to aliasing. 

\begin{figure*}[t!]%
    \centering
    \subfloat{{\includegraphics[width=13cm]{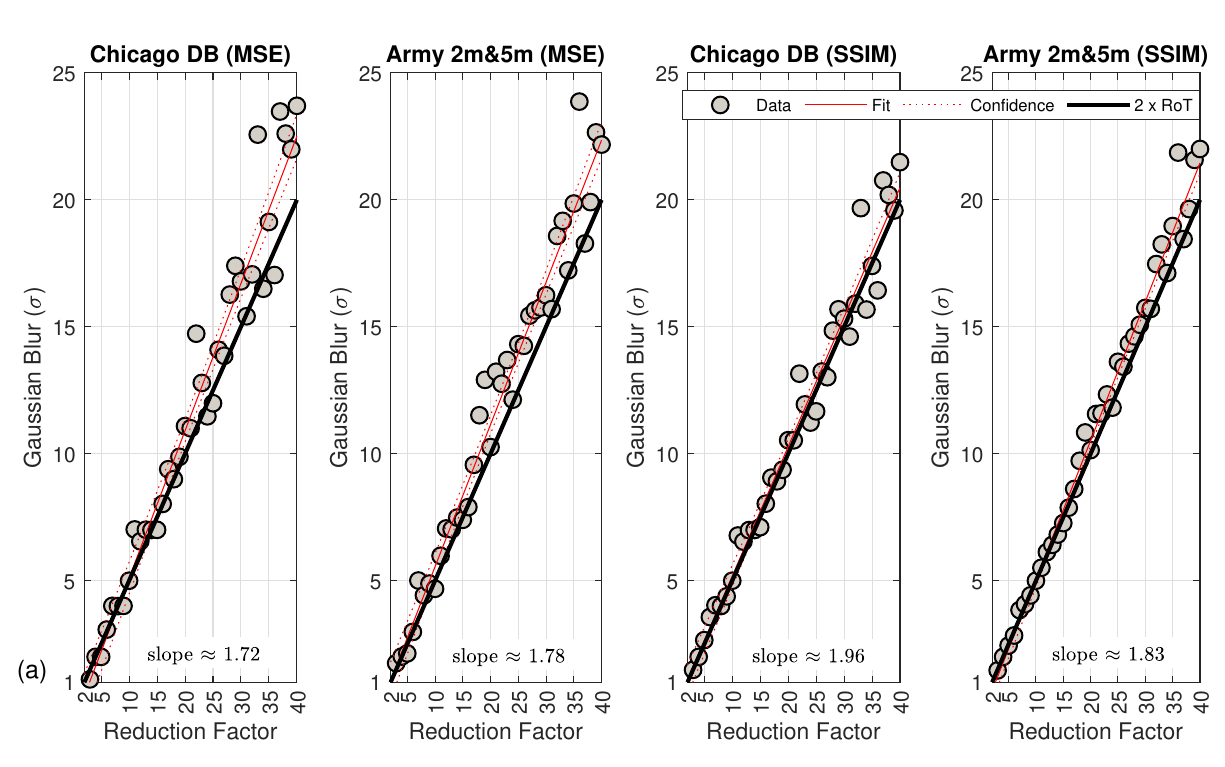} }}%
    \subfloat{{\includegraphics[width=4.8cm]{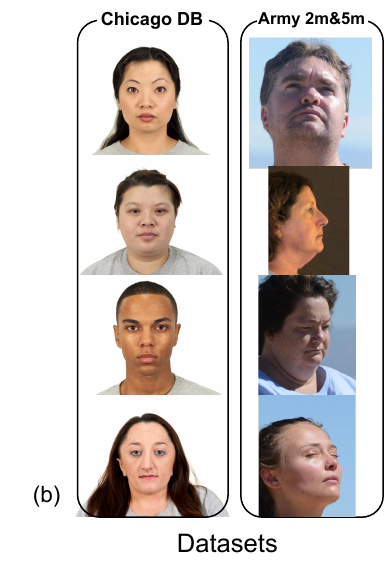} }}%
    \caption{(a) Empirically-observed relationship between the reduction factor and Gaussian Blur (kernel size expressed in terms of $\sigma$) for both datasets (Chicago DB and LRFID 2m\&5m) and both similarity measures (MSE and SIMM). The same figure also shows (dotted line) $95\%$ confidence intervals. (b) Depiction of examples from the two datasets used to conduct this experiment. RoT: Rule of Thumb.
    }
    \label{GAussianBlur}
\end{figure*}

Although the support of a Gaussian filter is infinite, we argued in the \textbf{Supplemental Materials} for small enough $m$\footnote{Note that for large $m$, $I_{dec}(w)$ would look like a noisy uniform signal in which case, the effect of interpolation filter (and its specially considered fall-off) would not matter anymore. The ultimate consequence of this observation is that the relationship we are trying to establish in our study will break apart.} that  $99.9\%$ of the area underneath (energy) sufficiently suppresses the high frequency and captures low-frequency components defined by the spectral decay of natural images in the frequency domain. To be able to get $99.9\%$ of the area underneath, it is sufficient to consider its support to be 3.291 of its standard deviation ($1/\sigma$ in the frequency domain due to the reciprocity property). In addition, one half of this support ($1.6455/\sigma$) corresponds to $90\%$ of the total area. Using this approximation, we consider two complementary scenarios.  
\begin{itemize}
    \item[] \textbf{Case 1:} \underline{There is no aliasing:} In this case, we guarantee that the compared images become practically identical due to no-aliasing and ideal filtering (no distortions).
    \item[]  \textbf{Case 2:} \underline{There is aliasing:} The optimal filtering requires to set the filter cut-off frequency in the intersection of the copies of the frequency response of the original signal i.e., precisely at $\pi/m$. Since our filters are not ideal (exhibiting true sharp transitions), the best we can do is to cover the entire signal spectrum with the frequency response of the filter i.e., just like in the no-aliasing case, leading to preserve $90\%$ of the original signal plus the $\approx 10\%$ of the aliased signal (image distortion). 
\end{itemize}
Consequently, considering both cases, our observation yields the following approximate relationship:
\begin{eqnarray}
    3.291 \times \frac{1}{\sigma} \approx  \frac{2 \pi}{m},
\end{eqnarray}
which leads to the relationship $m \approx 1.91 \sigma$. Note that this simplified analysis corroborates the approximate ``$2\times$" relationship between the standard deviation $\sigma$ of the isotropic
Gaussian kernel and the reduction factor $m$.

\begin{figure}[t!]%
    \centering
    \subfloat{{\includegraphics[width=9cm]{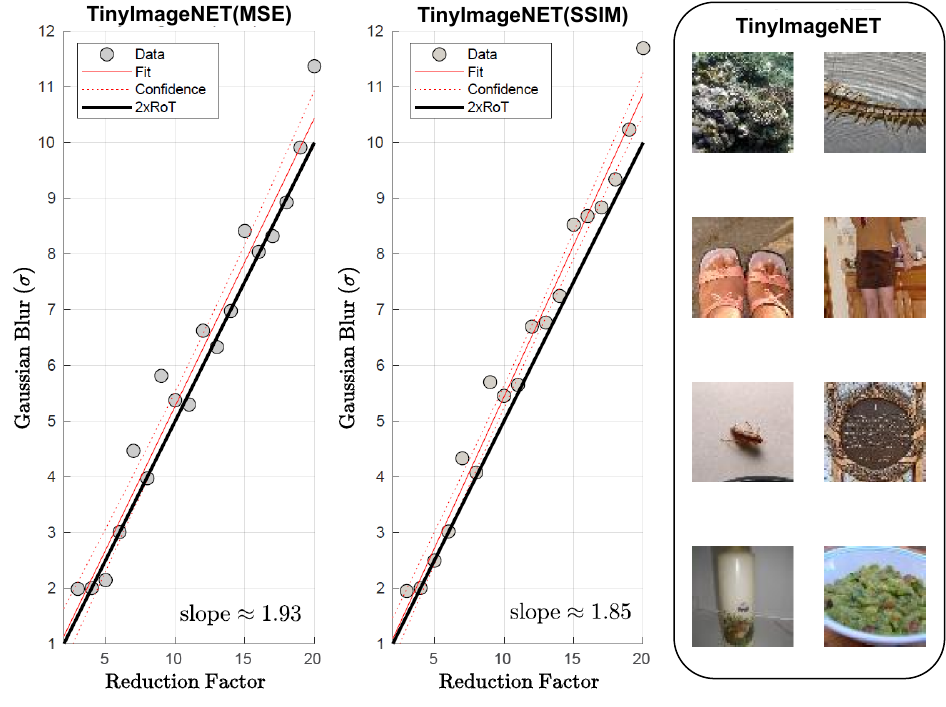} }}%
    \caption{Empirically-observed relationship between the reduction factor and Gaussian Blur (kernel size expressed in terms of $\sigma$) for TinyImageNET dataset, along with a few examples from the dataset. The same figure also shows (dotted line) $95\%$ confidence intervals. RoT: Rule of Thumb.}
    \label{GAussianBlur2}
\end{figure}

\section{Numerical Validation}

To support our theoretical arguments and experimentally establish the relationship between the standard deviation $\sigma$ and the reduction factor $m$, by adjusting the size of the Gaussian kernel and the rescaled image parameterized by $m$, we have run extensive tests for two different face datasets, namely Chicago DB \citep{ma2015chicago}  (with a total of 1203 clean images) and the LRFID dataset \citep{miller2019data} (with a total of 1350 clean images, primarily taken at distances 2m and 5m) having resolutions $2444 \times  1718$ and $400 \times 330 $, respectively. We have resampled these images with varying reduction factors ($m$) and investigated the means (as well as the confidence intervals) of the best Gaussian standard deviation $(\sigma^*)$ that minimizes the MSE or maximizes the SSIM $-$ using two different similarity metrics.  In Fig. \ref{GAussianBlur} (a), we have depicted the change of the parameter $(\sigma^*)$ as a function of the reduction factor, whereas Fig. \ref{GAussianBlur} (b) provides a few example faces from each database for visualization purposes.

We have expanded our findings to include natural scenes by leveraging the TinyImageNET dataset \citep{le2015tiny}, which comprises a grand total of 120,000 images, all with a smaller resolution of 64 $\times$ 64 pixels. These images have also been subjected to resampling at various levels of resolution reduction factors. Fig. \ref{GAussianBlur2} depicts the Gaussian standard deviation that minimizes MSE or maximizes SIMM, respectively, as a function of the resolution reduction factor $m$, along with a selection of sample images from the same dataset.

As can be seen in both sets of results, we clearly demonstrate that each resolution choice ended up with almost the same linear relationship between $\sigma$ and the reduction factor $m$. The slope values are observed to be close to our theoretical estimation of $1.9092$ ($p=0.056$ for $\%5$ confidence). Consequently, this relationship can be captured by a simple rule of thumb (RoT): Convolving an image with a Gaussian kernel results in resolution reduction (in each dimension) by a factor of approximately 2$\sigma$. The black bold line in the plots characterizes this ``rule of thumb" against the empirically determined regression fit -- red lines (and dotted line for confidence interval) for different resolutions. 
Furthermore, a notable observation emerges in terms of perceptual appropriateness for faces: the SSIM, as a similarity measure, exhibits superior alignment with the anticipated 2$\times$ rule of thumb, in comparison to the MSE. For natural images, the exact opposite can be observed, suggesting that while the MSE metric may excel in capturing fidelity for such images, it often fails to account for perceptual nuances critical in face processing and recognition tasks. 

\section{Conclusion}
In this letter, we have demonstrated a practical result that approximates an aliased image as a result of downscaling with the low spatial frequency content. More specifically, we have validated that the distortion caused by aliasing can be roughly replicated by applying a low-pass filter with an isotropic Gaussian kernel to the original image. Our findings  indicate a practical proportional relationship of approximately $2 \times$ between the standard deviation of the isotropic Gaussian kernel and the downsampling ratio used to downscale the image. This result could be particularly useful for simulating the effect of aliasing due to either distant viewing or degraded low-vision scenarios through remarkably straightforward low-pass filtering operations.

\section*{Acknowledgements}
This research is based upon work supported in part by the Office of the Director of National Intelligence (ODNI), Intelligence Advanced Research Projects Activity (IARPA), via [2022-21102100009]. The views and conclusions contained herein are those of the authors and should not be interpreted as necessarily representing the official policies, either expressed or implied, of ODNI, IARPA, or the U.S. Government. The U.S. Government is authorized to reproduce and distribute reprints for governmental purposes notwithstanding any copyright annotation therein.

\section*{Supplementary Material:  \\ Discussion on the area under Gaussian v.s. Spectral decay of Natural Images}

Natural images exhibit a quadratic power-law decay in their frequency spectra, often approximated as $1/f^2$ \citep{van1996modelling}(where $f$ is spatial frequency expressed in \textit{Hz}). In a log-log plot, this would mean that as frequency increases, the energy decreases linearly with a slope of about $–2$. In other words, most of the image’s energy is concentrated in the lower frequencies, and beyond a certain frequency (cut-off), the energy falls off very sharply. In fact, low frequencies (closer to 0) dominate the energy, whereas the high frequencies contribute minimally to total energy but encode fine details and edges.

A Gaussian function (either in time or frequency domain) is theoretically supported on the entire (infinite) real line, but in practice, we capture most of its energy within a finite window. As assumed in the main text that for $99.9\%$ of the area underneath (energy), the effective support is approximately $\pm 3.291 \sigma$. In contrast, using a threshold of, say $99.99\%$ would extend this to about $\pm3.719 \sigma$. Because of the reciprocity between the space and frequency domains, these spatial extents translate into corresponding effective cutoff frequencies in the frequency domain—roughly proportional to $1/ \sigma$ scaled by these constants.

Consider the derived relationship $3.291/\sigma \approx 2 \pi/m$ which, when solved for $m$, gives $m \approx 1.91 \sigma$, matching closely the $-2$ slope due to spectral decay of natural images. However, if we were to use, say, the 99.99\% area threshold $(3.719 \sigma)$, the relationship would become $m \approx (2\pi \sigma)/3.719 \approx 1.69 \sigma$. This noticeably lower factor would deviate from the experimentally observed linear trend between $m$ and $\sigma$.

Moreover, if you plot the logarithm of the effective area (or energy) versus $\log \sigma$, the slope we obtain is directly tied to this constant factor. The $99.9\%$ threshold yields a slope (or proportionality constant) that aligns well with the spectral decay of natural images, as the empirical data supports a nearly $2 \times$ relationship. On the other hand, other choices such as the 99.99\% threshold would yield a different slope, not matching the observed behavior as closely.

In addition to this reasoning, we also notice that the $99.9\%$ truncation strikes a balance between anti-aliasing and preserving perceptually relevant details in natural images. Over-suppressing high frequencies (e.g., with $99.99\%$ truncation) can lead to excessive blurring, degrading fine details that are perceptually important. More specifically, division of the frequency response of two Gaussians expressed as,
\begin{eqnarray}
\frac{G(\frac{\pi}{m},\sigma)}{G(\frac{2 \pi}{m},\sigma)} = e^{\frac{3\pi^2}{8}} 
\end{eqnarray}
which is $\approx 40 \times$ for $99.9\%$ truncation as opposed to $281 \times$ for $99.99\%$ truncation. This observation signifies a satisfactory trade-off that ensures the retainment of perceptually relevant details. 


\end{document}